\documentclass[10pt,twocolumn,prl,nofootinbib,preprintnumbers,amssymb,amsfonts,amsmath,superscriptaddress,showpacs]{revtex4-1}

\usepackage{graphicx}
\usepackage{bm}
\usepackage{hyperref}
\usepackage{color}
\usepackage{subfigure}
\usepackage{epstopdf}
\usepackage{wasysym}

\def\Fermi{{Fermi}}

\def\simleq{\; \raise0.3ex\hbox{$<$\kern-0.75em \raise-1.1ex\hbox{$\sim$}}\; }
\def\simgeq{\; \raise0.3ex\hbox{$>$\kern-0.75em \raise-1.1ex\hbox{$\sim$}}\; }
\newcommand{\eV}{{\rm eV}}

\newcommand{\GeV}{{\rm GeV}}
\newcommand{\TeV}{{\rm TeV}}

\newcommand{\erg}{{\rm erg}}

\newcommand{\kpc}{{\rm kpc}}
\newcommand{\pc}{{\rm pc}}

\newcommand{\cm}{{\rm cm}}
\newcommand{\km}{{\rm km}}

\newcommand{\s}{{\rm s}}
\newcommand{\sr}{{\rm sr}}

\definecolor{oucrimsonred}{rgb}{0.6, 0.0, 0.0}
\definecolor{persianblue}{rgb}{0.11, 0.22, 0.73}
\definecolor{forestgreen}{rgb}{0.13,0.35,0.13}
 \hypersetup{colorlinks, citecolor=oucrimsonred, linkcolor=persianblue, urlcolor=forestgreen}
 \hypersetup{colorlinks, citecolor=oucrimsonred, linkcolor=persianblue, urlcolor=forestgreen}

\begin{document}


\title{
Diffuse cosmic rays shining in the Galactic center: \\ A novel interpretation of H.E.S.S. and Fermi-LAT $\gamma$-ray data}


\author{D.~Gaggero}
\email{D.Gaggero@uva.nl}
\affiliation{GRAPPA, University of Amsterdam, Science Park 904, 1098 XH Amsterdam, Netherlands}

\author{D.~Grasso}
\email{dario.grasso@pi.infn.it}
\affiliation{INFN Pisa and Pisa University, Largo B. Pontecorvo 3, I-56127 Pisa, Italy}

\author{A.~Marinelli}
\email{antonio.marinelli@pi.infn.it}
\affiliation{INFN Pisa and Pisa University, Largo B. Pontecorvo 3, I-56127 Pisa, Italy}

\author{M.~Taoso}
\email{m.taoso@csic.es}
\affiliation{Instituto de F\'isica Te\'orica (IFT), UAM/CSIC,
Cantoblanco, Madrid, Spain}

\author{A.~Urbano}
\email{alfredo.leonardo.urbano@cern.ch}
\affiliation{CERN, Theoretical Physics Department, Geneva, Switzerland}

\begin{abstract}
We present a novel interpretation of the $\gamma$-ray diffuse emission measured by Fermi-LAT and H.E.S.S. in the Galactic center (GC) region and the Galactic ridge (GR). 
In the first part we perform a data-driven analysis based on {\tt PASS8} \Fermi-LAT data: We extend down to few GeV the spectra measured by H.E.S.S. and infer the primary cosmic-ray (CR) radial distribution between 0.1 and 3 TeV.
In the second part we adopt a CR transport model based on a position-dependent diffusion coefficient.
Such behavior reproduces the radial dependence of the CR spectral index recently inferred from the \Fermi-LAT observations.
We find that the bulk of the GR emission can be naturally explained by the interaction of the diffuse steady-state Galactic CR sea with the gas present in the Central Molecular Zone.
Although we confirm the presence of a residual radial-dependent emission associated with a central source, the relevance of the large-scale diffuse component prevents to claim a solid evidence of a GC Pevatron.
\end{abstract}

\maketitle

\section{Introduction}

The High Energy Stereoscopic System (H.E.S.S.) collaboration recently reported the discovery of a  $\gamma$-ray diffuse emission from a small region surrounding SgrA*  \cite{::2016dhk}. The emission spectrum is compatible with a single power-law with index $\Gamma_{\rm HESS16} = 2.32 \pm 0.05_{\rm stat} \pm 0.11_{\rm sys}$ and extends up to $\sim 50 ~\TeV$ with no statistically significant evidence of a cutoff.
If hadronic, as expected due to the strong losses suffered by electrons in that region, that emission may point to the presence of a proton population with energies up to the PeV in the Galactic center (GC). 

On the basis of the angular profile of the emission, the H.E.S.S. collaboration proposed the J1745-290 source as its possible origin. This source is positionally compatible with SgrA* supermassive black hole and with the G 359.95-0.04 pulsar wind nebula. Although the observed spectrum of HESS J1745-290 is suppressed above $\sim 10 ~\TeV$, this might be explained by the attenuation due to the presence of a dense radiation field around that source (see {\it e.g.} \cite{Celli:2016uon}).
Annihilating dark matter in the halo central spike \cite{Lacroix:2016qag}, or a peaked population of cosmic rays (CRs) interacting with high concentrated gas in that region, could also explain the diffuse emission measured by H.E.S.S.
The H.E.S.S. results have raised a wide interest as it seems to provide the first evidence of a Pevatron in our Galaxy.    
\begin{figure}[ht!]
\centering
\includegraphics[width=0.4\textwidth]{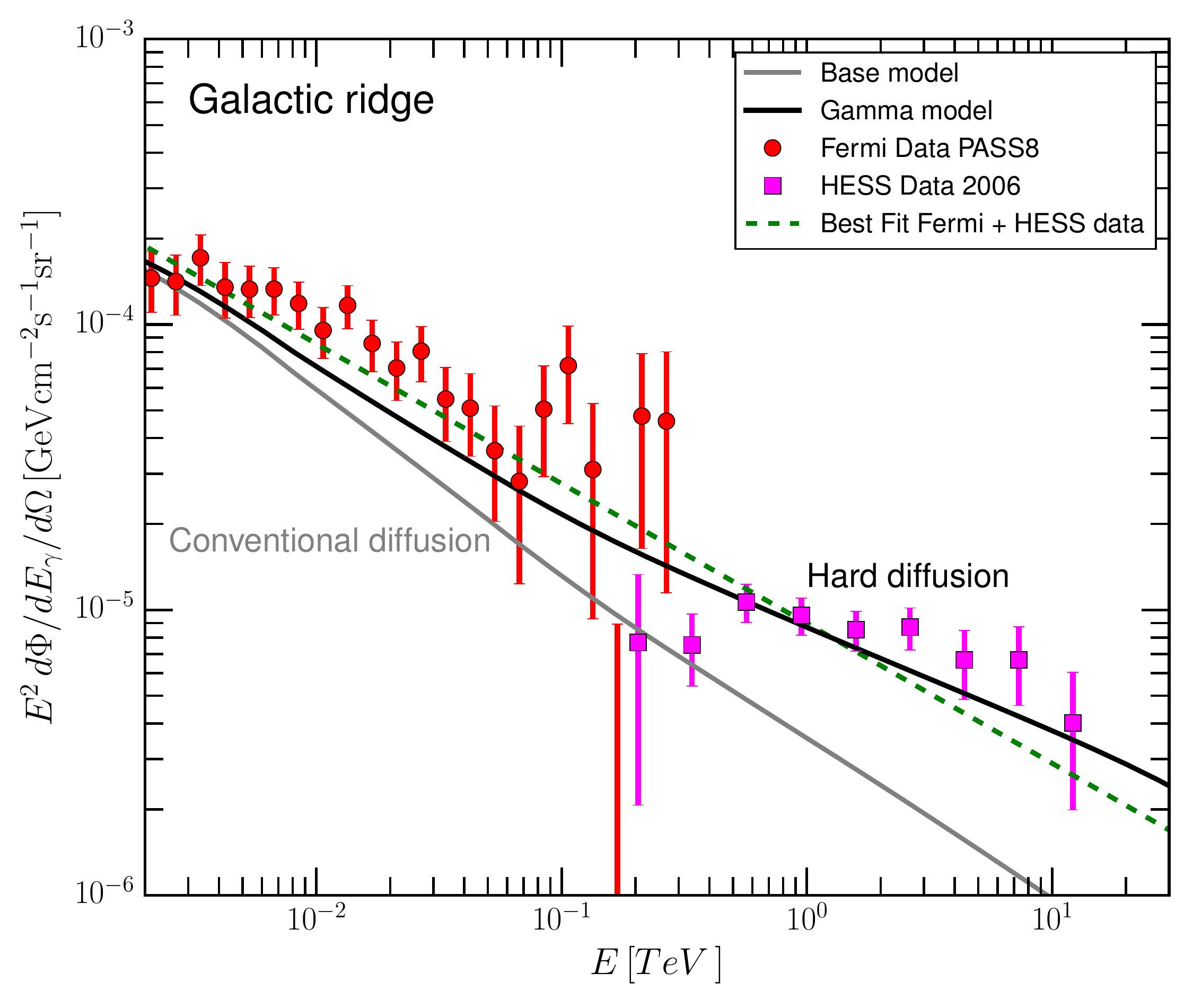}  
\caption{The $\gamma$-ray spectrum in the GR region ($| l | < 0.8^\circ$, $| b | < 0.3^\circ$). \Fermi-LAT data, shown here for the first time, and H.E.S.S. data from \cite{Aharonian:2006au} are compared with the contribution of the Galactic CR {\it sea} as computed with the {\it gamma} and {\it base} models discussed in the text. The single power-law best fit of the combined data is also reported. We have subtracted the contribution of point sources from \Fermi-LAT data.
}
\label{fig:spectrum_ridge}
\end{figure}

A $\gamma$-ray diffuse emission was also measured by a previous H.E.S.S. observational campaign towards the so called Galactic Ridge (GR) \cite{Aharonian:2006au}.  That emission approximately traces the gas distribution in the Central Molecular Zone (CMZ) -- a massive structure rich in molecular gas that extends up to $\sim 250~\pc$ away from the GC along the Galactic plane (GP). Its spectrum is compatible with a single power law with index $\Gamma_{\rm \rm HESS06} =  2.29 \pm 0.07_{\rm stat} \pm 0.20_{\rm sys}$, which, although observed only up to $\sim 10~\TeV$, is in agreement with that found in the inner region surrounding SgrA*.

The spectra of the CR population that one can infer from these data are significantly harder than the local CR spectrum measured at the Earth position ($\Gamma_{\rm CR}(r_\odot) \simeq  2.7$ for $E_{\rm CR} > 300~\GeV/{\rm nucleon}$ see {\it e.g.} \cite{Adriani:2011cu,Aguilar:2015ooa}). On the other hand, at lower energies, \Fermi-LAT observations of the SgrB complex in the CMZ suggest a CR spectrum similar to the local one~\cite{Yang:2014bcj}.

The H.E.S.S. collaboration proposed that the discrepancy could be the signature of a freshly accelerated CR population, possibly originated by SgrA* or by other sources in the central parsec of the Galaxy. 

The aim of this Letter is to estimate the contribution of the CR large-scale population (hereafter the CR {\it sea}) to the diffuse emission measured by H.E.S.S. and \Fermi-LAT in the GC region, and to provide a consistent interpretation of those data. 
Differently from previous computations, we model the CR {\it sea} by relaxing the simplified hypothesis of a uniform spectral index in the Galaxy.   
This approach is motivated by recent analyses of \Fermi-LAT data \cite{Gaggero:2014xla,Acero:2016qlg,Yang:2016jda} showing that the $\gamma$-ray diffuse emission of the Galaxy, and hence the CR primary spectrum, gets harder approaching the GC along the GP.  
  
In \cite{Gaggero:2014xla} this behavior was interpreted in terms of a radial dependence of both the scaling of the CR diffusion coefficient with rigidity, and the advection velocity.
The phenomenological model based on these ingredients reproduces the \Fermi-LAT data in most of the regions of the sky, including the inner GP where conventional models provide an unsatisfactory fit \cite{Ackermann:2012pya}. 
Later, it was shown \cite{Gaggero:2015xza} that the same scenario is in agreement with the high-energy data as well, providing a viable solution to the long-standing Milagro {\em anomaly}, i.e. an excess of the diffuse emission in the inner GP at $15$ TeV with respect to the predictions of conventional models \cite{Abdo:2008if}.  Moreover, this setting may also imply a significant Galactic contribution to the astrophysical neutrino flux recently measured by IceCube~\cite{Gaggero:2015xza} (see also \cite{Pagliaroli:2016lgg}).

Here we adopt the same scenario and, using a detailed 3D gas model for the CMZ region \cite{Ferriere:2007yq}, compute the contribution of the CR {\it sea} to the $\gamma$-ray diffuse emission from the GC region. 
We compare our results with 2006 and 2016 H.E.S.S. data and, for the first time in this context, with \Fermi-LAT {\tt PASS8} data for the same region.
We will show (see Fig. \ref{fig:spectrum_ridge} and \ref{fig:spectrum_pacman}) that -- above 10 GeV -- this contribution is significantly larger and harder than the one estimated so far on the basis of conventional models. 
Therefore we propose that a large fraction of the $\gamma$-ray emission measured by H.E.S.S. and \Fermi-LAT near the GC and in the whole GR is originated by the diffuse, steady-state Galactic CR {\it sea} interacting with the massive molecular clouds in the CMZ.

\section{Fermi-LAT data analysis}

\begin{figure}[ht!]
\centering
\includegraphics[width=0.4\textwidth]{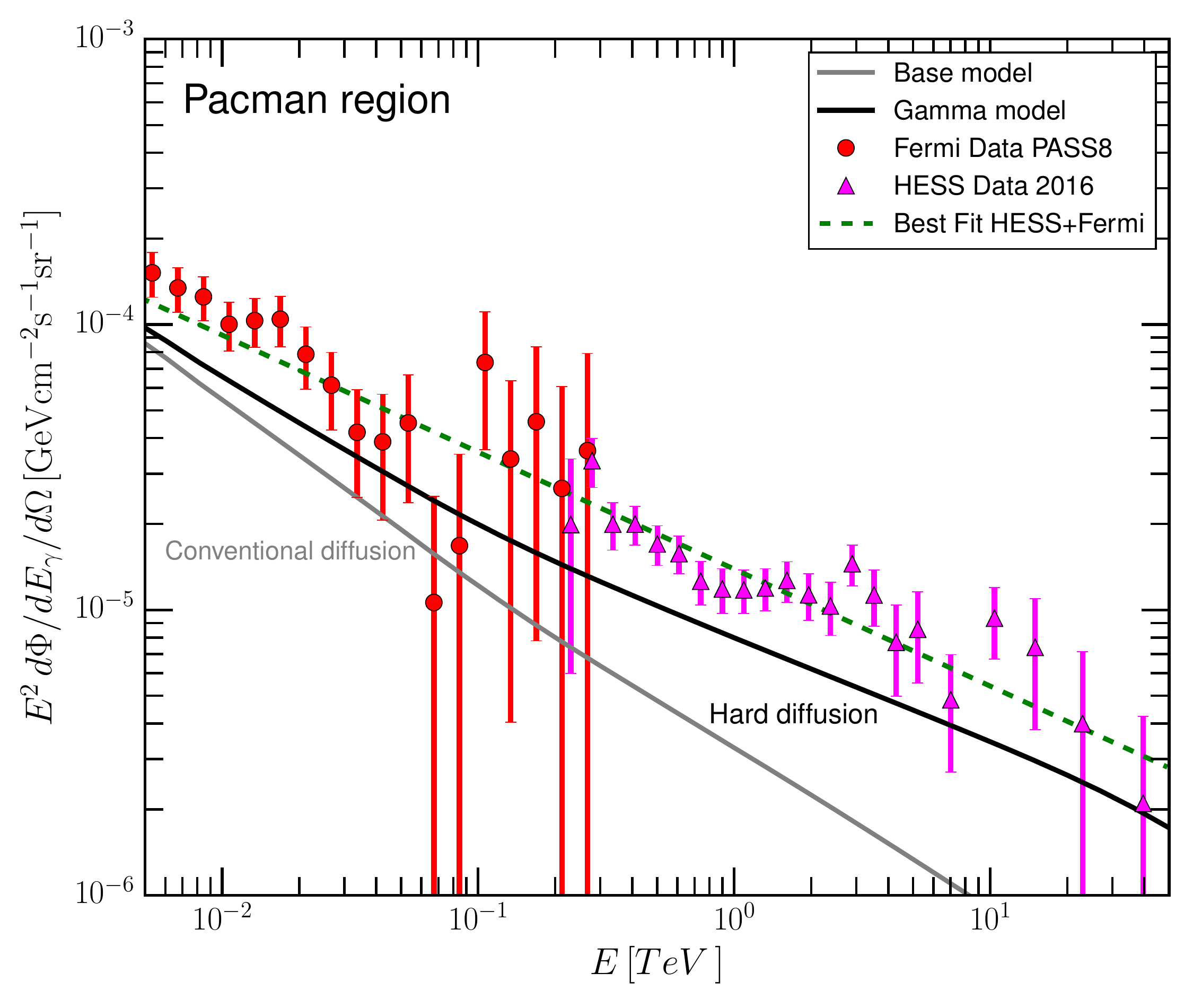}  
\caption{Same as Fig.~\ref{fig:spectrum_ridge} but for the {\it pacman} region defined in the text.
}
\label{fig:spectrum_pacman}
\end{figure}

The \Fermi-LAT collaboration recently released a new set of data based on the {\tt PASS8} event reconstruction algorithm \cite{Atwood:2013rka}.
In comparison to previous analyses, this approach yields a larger effective area, hence more statistics for the same observation time, as well as better energy and angular resolutions. 
Such improved performances are valuable in this context since they allow to improve the morphological and spectral information of the emission in the small portion of the sky under scrutiny.      

We extract \Fermi-LAT data using the \Fermi\ Science Tools {\tt v10r0p5} \cite{Ackermann:2012kna}. We use 422 weeks of {\tt PASS8} data with the event class {\tt CLEAN} and we apply the recommended quality cuts: {\tt (DATA\textunderscore QUAL==1) \&\& (LAT\textunderscore CONFIG==1)}. Moreover events with zenith angles larger than $90^\circ$ are excluded.
The exposure is computed using the \Fermi-LAT response function {\tt P8REP2\textunderscore CLEAN\textunderscore V6}.
The data are binned in 30 energy bins equally spaced in log scale between 300 MeV and 300 GeV.
The counts and the exposure maps have been produced using the {\tt HEALPix} pixelization scheme~\cite{Gorski:2004by}, with a resolution $n_{\rm side}=1024$, corresponding to a pixel size of $\sim 0.06^{\circ}.$

The emission from the point sources is obtained from the 4-year Point Source Catalog (3FGL) provided by the \Fermi-LAT collaboration~\cite{Acero:2015hja}. 
We also considered the high-energy 2FHL catalog finding only one source in the considered sky window, which is compatible with 3FGL J1745.6-2859c at the GC.
We model the point source emission convolving the flux of each 3FGL source with the point spread function (PSF) of the instrument, which is derived using the {\tt gtpsf} \Fermi  ~tool.

In  Fig.s \ref{fig:spectrum_ridge} and \ref{fig:spectrum_pacman} we report the \Fermi-LAT\ and H.E.S.S.
observations in the GR ($| l | < 0.8^\circ$, $| b | < 0.3^\circ$) and in the region considered in~\cite{::2016dhk}, an open annulus centered on SgrA* with $\theta_{\rm inner} = 0.15^\circ$  and $\theta_{\rm outer} = 0.45^\circ,$  (hereafter the ``{\em pacman}'').
The improved statistics provided by the {\tt PASS8} algorithm allows, for the first time, an overlap between Fermi-LAT and H.E.S.S. data around 200 GeV, covering therefore the entire energy range between $0.3$ GeV and 50 TeV.  
Noticeably, the two data sets are consistent with a single power law both in the GR and the {\em pacman} regions: The 95\% C.L. single-power-law fits from 10 GeV to 10 TeV in the two regions are respectively:
\begin{equation}
\Phi_{\rm{GR}} = 8.96^{+1.35}_{-1.39} \times 10^{-9} \left(\frac{E_{\gamma}}{1~\TeV}\right)^{{-2.49}^{+0.09}_{-0.08}}\left(\TeV~ \cm^{2}~ \s~ \sr \right)^{-1}
\end{equation}
and
\begin{equation}
\Phi_{\rm pm} = 1.36^{+0.12}_{-0.12}\times 10^{-8} \left(\frac{E_{\gamma}}{1~\TeV}\right)^{{-2.41}^{+0.07}_{-0.06}}\left(\TeV~ \cm^{2}~ \s~ \sr \right)^{-1}
\end{equation}
with reduced $\chi^2 = 3$, and 1.4.

We find only mild changes of our results using the \Fermi\ event type {\tt PSF3}, which corresponds to a subset of the events with a better angular reconstruction

In the rest of this section we use the angular dependence of the diffuse emission measured by \Fermi-LAT to infer the CR energy density radial profile $w_{\rm CR}(r)$ in the CMZ region, for energies corresponding to $E_{CR} \ge 100~\GeV$.  
We will then compare its shape with that determined by the H.E.S.S. collaboration for $E_{CR} > 10~\TeV$ \cite{::2016dhk}. Possible discrepancies among those profiles may reveal the presence of a non-stationary CR source since charged particles with different energies diffuse with different time scales. 
Moreover, \Fermi-LAT data extend to larger longitudes than H.E.S.S. which may allow to better probe the large radii tail of the CR distribution.   
  
For consistency, we determine $w_{\rm CR}$ using the same expression adopted in \cite{::2016dhk} (Eq. 2 in the Supplementary material of that paper) correcting it to account for the energy dependence of the pion production cross-section. This gives
\begin{eqnarray}
w_{\rm CR} &&(E_{\rm CR} \ge 0.1~\TeV) = 3.9 \times 10^{-2}~~\eV \cm^{-3} \nonumber\\
 &&\left(\frac{\eta_N}{1.5}\right)^{-1}~\left(\frac{L_\gamma(\ge 10~\GeV)}{10^{34}~\erg/\s}\right) \left(\frac{M_{\rm gas}}{10^6~M_\odot}\right)^{-1}~.
\end{eqnarray}
Here $L_\gamma(\ge E_\gamma)$ is the $\gamma$-ray luminosity above $E_\gamma$ in each region (subtracting the contribution from point sources); $M_{\rm gas}$ is the corresponding total hydrogen mass;
$\eta_N \approx 1.5$ is a factor accounting for the presence of heavier nuclei.   

\begin{figure}[th!]
\centering
\includegraphics[width=0.4\textwidth]{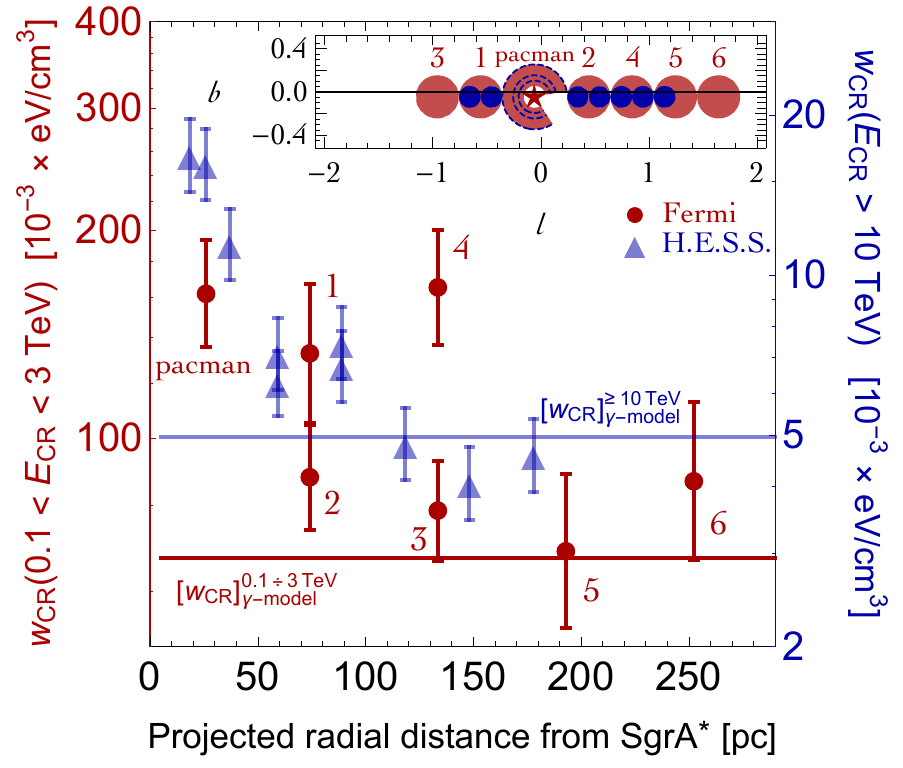}  
\caption{The CR energy density radial profiles for $E_{CR} > 10~\TeV$, as determined by H.E.S.S. \cite{::2016dhk}, and for $0.1 \le E_{CR} \le 3~\TeV$,  as determined here from \Fermi-LAT data, are reported. 
Those data are compared with the {\it gamma} model predictions (solid lines).
The regions of sky used for deriving the data are represented in the inset.
The model energy density profiles on Galactic scales are reported in the Supplementary material. 
}
\label{fig:cr_profile}
\end{figure}

Using the \Fermi\  tools we extract the diffuse luminosity $L_\gamma(E_\gamma \ge 10~\GeV)$ in an annulus and in six adjacent circular regions with angular diameter of $0.2^\circ$ centered on the plane intersecting SgrA* (see Fig.~\ref{fig:cr_profile}).  These regions are larger than those considered by H.E.S.S., which is motivated by the smaller angular resolution of \Fermi-LAT.
To determine the gas mass distribution we use the same CS column density map \cite{Tsuboi:1991} adopted by the H.E.S.S. collaboration \cite{::2016dhk}.

The resulting CR energy density radial profile $w_{\rm CR}(r)$ in the energy range $0.1 \le E_{CR} \le 0.3~\TeV$ is reported in Fig.~\ref{fig:cr_profile}, as well as the CR distribution derived by the H.E.S.S.  collaboration in \cite{::2016dhk} for $E_{CR} \ge 10~\TeV.$ 
Although the large errors and scatter of the points based on \Fermi-LAT data do not allow a tight constraint at low energies, our results are consistent with an energy independent shape of the CR density profile.
It is clear that both data sets are consistent with being constant for $r \simgeq 100~\pc$.


\section{Phenomenological Model }\label{sec:model}

In this section we compare the previous results with the phenomenological scenario proposed in \cite{Gaggero:2014xla}. 
This model was designed to reproduce the $\gamma$-ray spectra in the inner GP measured by \Fermi-LAT, which were found to be harder than those predicted by conventional models \cite{Ackermann:2012pya}. 
The scenario, which was implemented in the {\tt DRAGON} code \cite{Evoli:2008dv,Evoli:2016xgn}, assumes that the exponent $\delta$, setting the scaling of the CR diffusion coefficient with rigidity, has a linear dependence on the Galactocentric radius ($r$): $\delta(r) = A r + B$. The parameters $A$ and $B$ were tuned to consistently reproduce CR and Fermi-LAT $\gamma$-ray data on the whole sky. 
In particular the so-called KRA$_\gamma$ model adopts $A = 0.035~\kpc^{-1}$ and $B = 0.21$, giving $\delta(r_{\odot}) \simeq 0.5$. 
Assuming a uniform CR source spectral index across the whole Galaxy, this behavior turns into a radial dependence of the propagated CR spectral index, producing longitude-dependent $\gamma$-ray spectra along the GP.     
Remarkably, this is in reasonably good agreement with the results of a recent \Fermi-LAT  analysis \cite{Acero:2016qlg} (see Fig. 8 in that paper) as well as with those reported in \cite{Yang:2016jda} on the basis of the same data{\footnote{Note, however, that in \cite{Yang:2016jda} the CR spectrum at the GC is slightly softer than that found by the \Fermi-LAT collaboration}}. 
A radial dependence of the advection velocity was also adopted in \cite{Gaggero:2014xla}. Advection, however, plays no relevant role in the energy range considered in this work.

Similar to \cite{Gaggero:2015xza}, here we introduce a spectral hardening in the proton and Helium source spectra at $\sim 300~\GeV/{\rm n}$, in order to reproduce the local propagated spectra measured by PAMELA \cite{Adriani:2011cu}, AMS-02 \cite{Aguilar:2015ooa} and CREAM \cite{Ahn:2010gv}. 
We assume this feature to be present in the whole Galaxy, as it may be expected if it is produced by propagation effects. 
Under these conditions, the KRA$_\gamma$ model was shown \cite{Gaggero:2015xza} to reproduce the emission observed by Milagro in inner GP at a 15 TeV median energy \cite{Abdo:2008if} consistent with \Fermi-LAT data.  
At larger energies we assume a cutoff to be present in the CR source spectra at $\sim 5~{\rm PeV/nucleon}$, so as to match KASCADE-Grande results \cite{Apel:2013dga}, though this feature has no effect in the energy interval considered in this paper.

We compute the $\pi^0$, Inverse-Compton and bremsstrahlung components of the $\gamma$-ray diffuse emission, integrating the convolution of the spatially-dependent CR spectrum, gas/radiation density distributions and proper cross-sections along the line-of-sight. The $\pi^0$ component is dominant in the GC region. 
We checked that, for reasonable choices of the interstellar radiation field (ISRF), the $\gamma$-ray opacity in the CMZ region is negligible in the energy range considered in this work. Here we adopt the ISRF taken from the latest public version of {\tt GALPROP} \cite{Galpropweb,Vladimirov:2010aq} and pion production cross sections as parametrized in \cite{Kamae:2006bf}.
The scenario proposed in \cite{Gaggero:2014xla} predicts a CR proton spectral index $\Gamma_{\rm CR}(r \simeq 0) = \Gamma_{\rm CR}(r_\odot) - A r_\odot$.  Then, since $\Gamma_{\rm CR}(r_\odot) \simeq 2.7$ above $\sim 300~\GeV$, this implies $\Gamma_{\rm CR}(r \simeq 0) \simeq 2.4$, in agreement with what found in the previous Section. 
We notice that this finding is independent on the value of the parameter $B$ setting the normalization of $\delta$ so that tuning this quantity to better match recent AMS-02 B/C results \cite{Aguilar:2016vqr} would not affect our results.

With respect to what reported in \cite{Gaggero:2014xla}, here we replace the hydrogen distribution in the inner $3~\kpc$ with the 3-dimensional analytical model presented in \cite{Ferriere:2007yq}, as required to properly model the hadronic emission in that region.  Outside that region we adopt the gas model used in \cite{Galpropweb,Vladimirov:2010aq}.  
The main components are molecular (H$_2$) and atomic (HI) and hydrogen. HI, which is inferred from 21-cm lines, is less than 10\% of the total mass. 
H$_2$ is not observed directly; except for the densest clumps, the column density can be inferred from several tracers, most commonly from the CO emission lines. This requires a conversion factor which was estimated to be $X_{\rm CO}(r \sim 0) \simeq 0.5 \times 10^{20}~\cm^{-2}~{\rm K}^{-1}~\km^{-1} \s$ with a factor 2 uncertainty \cite{Ferriere:2007yq}.  
Here we use $X_{\rm CO}(r \sim 0) \simeq 0.6 \times 10^{20}~\cm^{-2}~{\rm K}^{-1}~\km^{-1} \s,$ the value giving the best agreement with the integrated mass distribution, based on the CS emission map,  reported in \cite{::2016dhk}. 
The quoted uncertainty on this parameter directly applies to the diffuse $\gamma$-ray emission normalization. 
This effect, however, is degenerate with that of varying the CR (poorly known) source density at the GC.  

Following \cite{Gaggero:2014xla}, we use the CR source distribution of~ \cite{Case:1998qg}, based on supernova remnant catalogs.
This parametrization vanishes at the GC, a behavior in qualitative agreement with the $\gamma$-ray emissivity profile determined by the \Fermi-LAT collaboration \cite{Acero:2016qlg}, which displays a dip in the GC.
For a given transport model, this choice minimizes the CR {\it sea} density in the CMZ region.  
We verified that using the source distribution reported in \cite{Yusifov:2004fr}, which does not vanish at the GC, turns into a factor $\sim 2$ larger emission from the GR and {\it pacman} regions. This is still compatible with experimental data. Moreover, this enhancement can be compensated by a reduction of the $X_{\rm CO}$ factor leaving it well within the observationally allowed range.

In Fig.s \ref{fig:spectrum_ridge},\ref{fig:spectrum_pacman} we show the gamma-ray emission in the GR and {\it pacman} regions predicted by this model (hereafter {\it gamma model}).
For comparison we also report the spectrum computed for a conventional model ({\it base model}), sharing with the {\it gamma model} all the properties but keeping the diffusion coefficient and the convective velocity spatially uniform {\footnote{The main parameters characterizing the {\it base} and {\it gamma models} are reported in Tab. 1 in the Supplementary material (SP).}}
We find that the {\it base} (as any other conventional) {\it model} cannot consistently reproduce  the H.E.S.S. and \Fermi-LAT measurements in the absence of an additional component with a harder spectrum.  Instead, in the GR the {\it gamma model} is in excellent agreement with the shape and normalization of the measured spectrum. 

In the {\it pacman} region the {\it gamma model} prediction lies slightly below the data.
This is consistent with what inferred from the CR energy density radial profile $w_{\rm CR}$, shown in Fig. \ref{fig:cr_profile}, which respect to the CR sea (almost uniform on those small scales) 
displays a peak toward the GC.  We interpret this feature to be due to one (or more) CR source(s) in the inner few pc of the Galaxy.
Far outside that region ($r \simgeq 100~\pc$), we find that the CR energy density in both energy ranges $0.1 \le E_{CR} \le 0.3~\TeV$ and $E_{CR} \ge 10~\TeV$, is in good agreement with experimental data.
 
Although not showed here, we have also checked that the {\it gamma model} is in excellent agreement with the \Fermi-LAT and H.E.S.S. observations in the SgrB complex region ($0.4 < l < 0.9$, $-0.3 < l < 0.2$). 
 
\section{Conclusions}

In this Letter we have shown that a large fraction of the $\gamma$-ray emission from the CMZ measured by H.E.S.S. and \Fermi-LAT from few GeV up to 50 TeV can be originated by the same population of energetic particles.
In fact, we have found that the Galactic CR {\it sea} accounts for the bulk of that emission if it is modeled under the assumption of a spatial-dependent diffusion. This feature is motivated by the radial dependence of the CR spectral index recently inferred from \Fermi-LAT data. 
Therefore, our results provide a new strong evidence supporting the validity of that scenario
in a region of the Galaxy were the discrepancies between the base and conventional model are expected to be maximal.
 
On top of this diffuse emission, we have outlined -- by means of an energy-dependent data-driven analysis -- the hint for the presence of a larger CR density in the vicinity of the central radio source Sgr A* with respect to the average density inferred from the whole GR, similarly to that what found by H.E.S.S. at larger energy. We have not found any significant evidence of a different spectral shape between those regions. 

Therefore, this excess may be originated by one or more CR  sources in the inner few parsecs of the Galaxy which are likely to be responsible for the J1745-290 emission.
No firm conclusion, however, can be drawn on the maximal energy of CR accelerated by these sources since the significance of the $\gamma$-ray excess in the {\it pacman} region with respect to the contribution of the CR sea is rather small above 10 TeV. 
 
In future, the South site of CTA \cite{CTAofficial} may provide a further confirmation of the scenario discussed in this Letter from the detailed observation of a larger region centered on the GC.

\vskip0.5 cm

\noindent{\underline{\em Acknowledgments}:}
We warmly thank L.~ Baldini, C.~Evoli, S.~Gabici, M.~Pesce-Rollins, P.~Ullio for valuable discussions. 
We also thank A.~Viana for providing us the CS column density map used by the H.E.S.S. collaboration.
M.T. is supported by the centro de Excelencia Severo Ochoa Programme SEV-2012-0249, MINECO through a Severo Ochoa fellowship with the Program SEV-2012-0249, FPA2015-65929-P and Consolider MultiDark CSD2009-00064.
Pac-Man was designed by T.Iwatani.


\bibliographystyle{apsrev4-1}
\bibliography{bibpevatron}

\end{document}